\documentclass[sigconf]{acmart}

\pdfoutput=1

\usepackage{booktabs} 

\usepackage{wrapfig}
\usepackage{ifthen}
\usepackage{algorithm}
\usepackage[noend]{algpseudocode}
\usepackage{makecell}

\acmConference[DEBS 2017]{The 11th ACM International Conference on Distributed and Event-Based Systems}{June 2017}{Barcelona, Spain}

\newboolean{showcomments}
\setboolean{showcomments}{false}
\ifthenelse{\boolean{showcomments}}
{ \newcommand{\mynote}[3]{
    \fbox{\bfseries\sffamily\scriptsize#1}
    {\small$\blacktriangleright$\textsf{\emph{\color{#3}{#2}}}$\blacktriangleleft$}}}
{ \newcommand{\mynote}[3]{}}

\newcommand{\eo}[1]{\mynote{Emi}{#1}{green}}

\begin{document}
\title{Grand Challenge: Optimized Stage Processing for Anomaly Detection on Numerical Data Streams}
\renewcommand{\shorttitle}{Optimized Stage Processing for Anomaly Detection on Numerical Data Streams}

\author{Ciprian Amariei}
\affiliation{%
  \institution{Alexandru Ioan Cuza University of Ia\c{s}i}
  \country{Romania}
}
\email{ciprian.amariei@gmail.com}

\author{Paul Diac}
\affiliation{%
  \institution{Alexandru Ioan Cuza University of Ia\c{s}i}
  \country{Romania}
}
\email{paul.diac@info.uaic.ro}

\author{Emanuel Onica}
\affiliation{%
  \institution{Alexandru Ioan Cuza University of Ia\c{s}i}
  \country{Romania}}
\email{eonica@info.uaic.ro}


\begin{abstract}
The 2017 Grand Challenge focused on the problem of automatic detection of anomalies for manufacturing equipment. This paper reports the technical details of a solution focused on particular optimizations of the processing stages. These included customized input parsing, fine tuning of a k-means clustering algorithm and probability analysis using a lazy flavor of a Markov chain. We have observed in our custom implementation that carefully tweaking these processing stages at single node level by leveraging various data stream characteristics can yield good performance results. We start the paper with several observations concerning the input data stream, following with our solution description with details on particular optimizations, and we conclude with evaluation and a discussion of obtained results.
\end{abstract}



\copyrightyear{2017}
\acmYear{2017}
\setcopyright{acmcopyright}
\acmConference{DEBS '17}{June 19-23, 2017}{Barcelona, Spain}\acmPrice{15.00}\acmDOI{10.1145/3093742.3095101}
\acmISBN{978-1-4503-5065-5/17/06}

\maketitle

\section{Introduction}
\label{sec:intro}

The 2017 DEBS Grand Challenge focused on a use case of analyzing data streams generated by sensors embedded in manufacturing equipment with the goal of detecting anomalies in the equipment behavior.
The query to be solved for the anomaly detection involved three main processing stages to be executed for each sensor: (1) finding clusters of sensor measurements, (2) training a Markov model for detecting the probability that sensor reports shift between clusters and finally (3) detecting anomalies based on the frequency of shifting reports correlated with the trained Markov model.
In addition, our solution included a preprocessing stage of the data, which was provided as a time-ordered stream of RDF tuples.
A full description of the proposed problem is thoroughly detailed in~\cite{DEBS2017} \eo{to change reference to the one they provide}.

The solution we propose in the current paper focuses on optimizing the processing chain at the level of a single node by leveraging several particularities we observed in both the structure of data stream and in the problem characteristics.
The data to be analyzed was embedded in an RDF format with a fixed structure. 
Each message received as input represented an observation group associated with a single machine and contained all sensor measurements for that respective machine at a certain time.
We have observed that the fixed RDF structure permitted a fast customized parsing as part of the preprocessing stage, which allowed us to easily create dedicated processing queues for each individual sensor. \eo{refer some more exact characteristics of the ontology here?}

Further, the target query composed of the aforementioned three stages should be solved with respect to a sliding time window over each of the created sensor queues.
We have observed that for some windows the result of some processing stages can be reused, based on computation executed over previous windows.
This allows us to skip the query computation in such cases, which we noticed to have a relatively high frequency, and to rely on previously obtained results.
This optimization impacted both our K-means implementation for finding the clusters in the first stage, as well as the second stage training of the Markov model.
We also considered additional optimizations for these two stages, which reduce the complexity of the baseline algorithms.

In our implementation, we also leverage the fact that the received observation groups can be ordered based on timestamp as well as the fact that processing of queues for different sensors does not interfere in respect with the result.
This allowed us to implement a simple parallelization at single node level of our processing chain.
In this we just divide the work to multiple threads, where a particular sensor queue will always be handled by the same thread. \eo{is it a machine or a sensor? and if it is a machine, why couldn't be a sensor?...}
We believe that our baseline design for the single node solution is further extendable to a multi-node implementation with minor adjustments and added synchronization in the final step.
However, based on preliminary runs using a multiple node configuration, we concluded that under the given conditions of the test platform, a distributed solution would not perform significantly better (or actually might have even worse results) than the baseline approach.

We structure our paper as follows.
In Section~\ref{sec:solution} we describe our solution architecture including technical implementation details.
In Section~\ref{sec:optimizations} we discuss the particular optimizations we have implemented and their impact.
We continue in Section~\ref{sec:evaluation} with measurements obtained on the test platform.
We conclude in Section~\ref{sec:discussion} with further ideas of optimization and with a discussion on the obtained results.


\section{Solution Architecture}
\label{sec:solution}

We developed our solution in the Java language, using a custom pipelined architecture.
The reason for choosing Java was based mostly on the fact that the testing platform was written in Java, involving several standard wrapper classes facilitating the integration of the solution with the benchmark.
The rationale of preferring a custom architecture to a dedicated stream processing platform (e.g., Storm~\cite{AS2011}) 
had mainly two grounds.
The first was due to some optimization cases which are selectively triggered based on the window values configuration and which involve all three processing stages, making improper to adhere to the specific operator-like separation that stream processing platforms typically imply.
The second ground was the single node approach we finally decided for.

The first step of our solution consists in loading the metadata describing the problem parameters, performing the necessary initialization of data structures and threads, and also executing a warmup phase.
In particular, one dispatcher thread is created for partitioning the workload, and a set of worker threads for executing the processing stages.
We preferred using, where possible, either singleton or static versions for the necessary data structures in order to avoid costs of reinstantiating.
During the warmup phase, we are running the entire detection pipeline using 5000 observation groups for 3 times (3 was determined as the best option in respect to overall results after a couple of trials with various values).

\begin{figure*}
\includegraphics[width=\textwidth]{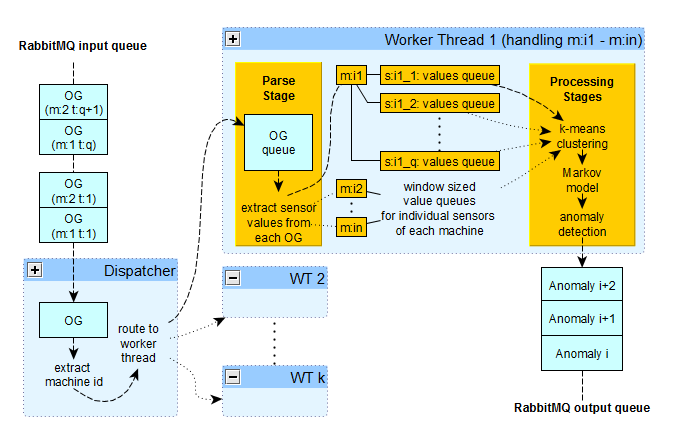}
\centering
\caption{Pipeline of processing an observation group. Notations: $m:i$ - machine with identifier i ; $t:i$ - timpestamp i ; $s:i\_j$ - values of sensor i on machine j.}
\label{fig:arch}
\end{figure*}

We present our custom pipeline architecture in Figure~\ref{fig:arch} following the path of an observation group starting with receiving it from the input queue up to generating an anomaly.

An observation group is read from the RabbitMQ input queue as a vector of bytes in RDF format by a dispatcher thread.
This thread extracts the machine id from the observation group with a minimum of effort, by simply searching for it starting at a very specific position in the byte array.
This position is determined by counting the length of the starting part of the RDF entry, which is similar for each observation group because it follows a specific pattern according to the given ontology.
Proceeding like this minimizes the search effort in the given input.
This is essentially performed in $O(1)$ with a minor constant given by the text length corresponding to ids increasing in time, which is variable.

Further, the observation group is routed by the dispatcher thread to an internal preprocessing queue maintained by each worker thread.
Currently, the dispatching algorithm follows just a simple routine, by applying a modulo on the machine id to identify the thread chosen for processing.
One important invariant guaranteed by this strategy is that all observation groups generated by the same machine are always processed by the same thread in the order they are received (e.g., in Figure~\ref{fig:arch} the first worker thread will always and exclusively process the data generated by the machines in the group $m:i1$ to $m:in$).
As we observed the input is currently received sequentially as one observation group per each machine.
Therefore, this strategy works efficiently as long as all machines are producing data.
In the case of machines leaving and joining, the routing would obviously require a different strategy for maintaining a proper load balancing.

As illustrated in Figure~\ref{fig:arch} a worker thread will further execute both the initial parsing stage and the chain of processing stages over each observation group received in its queue.
First, the worker thread will apply a similar technique as previously mentioned for efficiently parsing and extracting a sensor value from the observation group RDF data.
The thread will essentially skip fixed constant portions of RDF data and start searching for a sensor property id and associated sensor value at specific positions, typically immediately close to the actual searched data location.
Since this parsing is done essentially in $O(1)$ time for each sensor value and since different sensor values do not influence each other in the processing, the thread will continue sending each obtained value through the processing pipeline before extracting the next value.
The position that the parsing reached in the RDF observation group data is obviously saved for continuing after the current obtained sensor value is processed.

Each sensor value obtained by the parsing is added to a window sized queue associated with that specific sensor, which essentially represents the current window of values to be processed (e.g., $s:i1\_1$ for the example in Figure~\ref{fig:arch}).
All these queues are organized in a double indexed array, where the first index identifies the machine id and the second index the sensor property id.
This structure, available at the global level, provides easy access without any race conditions and in constant time to the individual window queues by the worker threads.
Whenever a new value is added to a queue the first one is removed, sliding the window.
Afterwards, the thread starts executing the chain of processing stages over the values present in the window queue (e.g., over $s:i1\_1$ in the example in Figure~\ref{fig:arch}).

In the processing stages, we have applied several optimizations, some of which depending on the specific structure of the windows with respect to the contained values.
We detail our implementation of analyzing the values in a given window and detecting an anomaly in Section~\ref{sec:optimizations}.
After an anomaly is produced by a worker thread it is forwarded to a pre-exit priority queue ordered according to the current timestamp.

An anomaly resides in this queue for a short time, until it is determined that no worker thread could produce another anomaly with an earlier timestamp.

In order to determine that no threads could produce an earlier anomaly, each worker thread signals its last processed timestamp to a central non-blocking data structure.
A separate thread is used to periodically check the latest minimum processed timestamp.
Once all threads reached a higher processed timestamp, all queued anomalies are flushed to the RabbitMQ output queue as long as their timestamp is lower than the minimum of already processed timestamps.

In the case when the number of machines sending sensor values is not constant, one or more threads could enter a starvation phase.
If all machines assigned (via the routing process) to the same thread go silent, the thread would have an empty message queue.
This blocks the output queue as its latest processed timestamp does not increase anymore until its machines start sending data again.
In order to prevent this situation, the starving thread would signal with a fixed maximum timestamp value so it would not count when computing the minimum of the latest processed messages.

\section{Processing Chain Optimizations}
\label{sec:optimizations}

As described in Section~\ref{sec:solution} the chain of processing stages is executed by each worker thread in sequence for every sensor value that is parsed and added to the window queue.
Besides the processed window queue we particularly maintain during the entire chain of processing several data items used by the optimizations we implemented.
These items are:

\begin{itemize}
\item \emph{prev\_first} - the first value of the previous window that was just removed from the current window;
\item \emph{frequencies} - the frequency of apparition of the first $K$ distinct values in the window, up to the first apparition of the $K^{th}$ distinct value, where $K$ is the number of clusters used by the K-means algorithm;
\item \emph{position} - the position of the last distinct value determined in the previous window;
\item \emph{cluster\_sequence} - the sequence of clusters associated with each value in the window after the iteration of the K-means stage.
\end{itemize}

We detail in the following the optimizations implemented in each stage, some of which span across all stages.

\subsubsection*{K-means clustering.}

\begin{figure}
\includegraphics[width=0.45\textwidth]{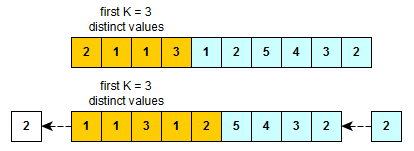}
\centering
\caption{Window sliding case when clusters remain unchanged - the case of \emph{IN/OUT} optimization.}
\label{fig:kmeansopti}
\end{figure}

The first optimization in this stage, which we further refer as~\emph{IN/OUT}, exploits the situation when the last value added to the processed window queue is equal to \emph{prev\_first} and the first $K$ distinct values in the window remain the same.
For clarity we illustrate an example in Figure~\ref{fig:kmeansopti} where \emph{prev\_first}=2.
The challenge specification requires setting the initial cluster centers to the first $K$ distinct values in the window.
We can notice that in such context the clusters identified after the K-means stage will be the same as in the previous window.
This is due to the fact that we start the clustering having the same $K$ distinct initial centers and the same values in the new window.
It is, therefore, enough to identify this situation in order to skip the K-means clustering.
The Algorithm~\ref{alg:inout} we implemented to determine this context runs in sublinear time with respect to the window size.
We describe the algorithm in the following.

\begin{algorithm*}[t]
\caption{IN/OUT algorithm}\label{alg:inout}
\begin{algorithmic}[1]
\Function{in/out}{}
\State $result\gets false$
\State $frequencies[prev\_first]--$\label{lin:1} 

\If{$frequencies[prev\_first] == 0$}\label{lin:2} 
\State $frequencies.remove(prev\_first)$\label{lin:3} 
\While{$position < window.length \land frequencies.size() < K \land frequencies.includes(window[position])$}\label{lin:4} 
    \State $frequencies[window[position]]++$\label{lin:5} 
    \State $position++$\label{lin:6} 
  \EndWhile

  \If{$position < window.length$}\label{lin:7} 
      \State $frequencies.add(window[position])$ \label{lin:8} 
    \State $frequencies[window[position]] \gets 1$\label{lin:9} 

    \If{$window[position] == prev\_first$}\label{lin:10} 
      \State $result\gets true$\label{lin:11} 
    \EndIf
  \Else
      \State $position--$\label{lin:12} 
  \EndIf
\Else
\label{lin:13} 
\State $result\gets true$\label{lin:14} 
\State $position--$\label{lin:15} 
\EndIf
\State\Return $result \land (prev\_first == current\_last)$
\EndFunction
\end{algorithmic}
\end{algorithm*}

We first decrease at line 3 the maintained frequency for \emph{prev\_first} since one such value left the window.
At line 4, we check if there are no other similar values to \emph{prev\_first} in the range of first distinct $K$ values determined in the previous window.
If there are other similar values in this range as \emph{prev\_first} then we know that the window cluster centers are the same (line 17) and the algorithm ends by checking also the condition of the last value added being equal to \emph{prev\_first}.
If there are no other values equal to \emph{prev\_first} in the window we first remove the corresponding value key from the \emph{frequencies} map (line 5).
This means, in the general case, that we are one distinct value less then the number of clusters.
Following, at line 6 we try to find if there is another distinct value in the window, starting at the position of the last known distinct value obtained in the previous window.
If we find such a distinct value by the end of the window (line 9) then we add it to the frequency map and we check if it is equal to \emph{prev\_first}.
If that is the case we know again that we have the same first $K$ distinct values as the previous window.
The other decrements of \emph{position} relate with keeping this data valid for future windows.

Using this technique the standard complexity of the K-means stage is reduced from $O(K \times W)$ to $O(W)$ where $W$ is the window size.
Furthermore,~\emph{IN/OUT} has also implications in the Markov modelling stage, as we will describe below.

The second optimization implemented in this stage, which we will refer to as~\emph{K1}, is the situation where the values in a given window will be clustered in a single cluster.
This can happen either in the case when the metadata defines only 1 cluster for a given sensor, or when all the values in the window are identical.
The~\emph{K1} situation permits skipping not only the K-means processing stage but all the processing chain.
Since all values are part of the same cluster, the probability of transition for each value will be 1 and there will be no anomalies detected in the given window.
The complexity of the entire processing chain is therefore reduced to $O(W)$ necessary to determine if all values are equal (except the case when the metadata defines 1 cluster for the sensor when the complexity is $O(1)$).

The third optimization in this stage, which we refer further as~\emph{LowK}, refers to the case when the number of distinct values in the window is lower than the number of clusters $K$ defined in the metadata for the respective sensor.
This permits again skipping the complete K-means clustering stage since we can directly assign each value to a corresponding cluster knowing that values will not shift between clusters if we execute any other iteration.
The~\emph{LowK} optimization reduces again the complexity of K-means from $O(K \times W)$ to $O(W)$ necessary for iterating through the window for assigning the values.

As we detail in Section~\ref{sec:evaluation} we observed that the occurrence of the aforementioned situations in the data sets offered for testing is considerably high, the optimizations being, therefore, triggered very often.

\subsubsection*{Markov modelling.}

In the Markov modelling stage, we make use of the \emph{cluster\_sequence} determined in the K-means clustering.
Using this we need to compute the probability of transition from one cluster to another.
The probability of transition from cluster $a$ to cluster $b$ is defined as the observed number of transitions from $a$ to $b$ in the window, divided by the total number of transitions from $a$ observed in the window.
We decided for a lazy approach in our implementation, where we only count in this stage the two numbers that we need to divide.
This results in $O(W)$ time complexity for this stage. 
The optimization helps skipping another $O(K^2)$ operations that the division between each possible cluster transition count would require.
Moreover, in the case where the K-means~\emph{IN/OUT} optimization applies, we can observe that the values in the \emph{cluster\_sequence} simply shift, the former first one becoming the last.
In this situation, we can skip the full count execution in the Markov modelling stage.
This is done in $O(1)$ by just preserving the counted values from the previous window and updating the corresponding counts for the first and last value in the \emph{cluster\_sequence}.

\subsubsection*{Anomaly detection.}

The lazy implementation of the Markov modelling stage implies that we need to finalize computing the actual probabilities of transition in the anomaly detection stage.
This is done by dividing the transition counts obtained previously.
However, instead of computing the entire $K^2$ set of probabilities for all the possible cluster transitions, we only compute the probabilities for the last $N=5$ transitions as in the given specification of~\cite{DEBS2017} (which can be even less if the last values in the window did not shift between clusters).
Finally, we detect an anomaly by calculating the composed probability of these last transitions and comparing it with the given reference threshold.

\section{Evaluation and Observations}
\label{sec:evaluation}

\begin{figure*}[th!]
\includegraphics[width=\textwidth]{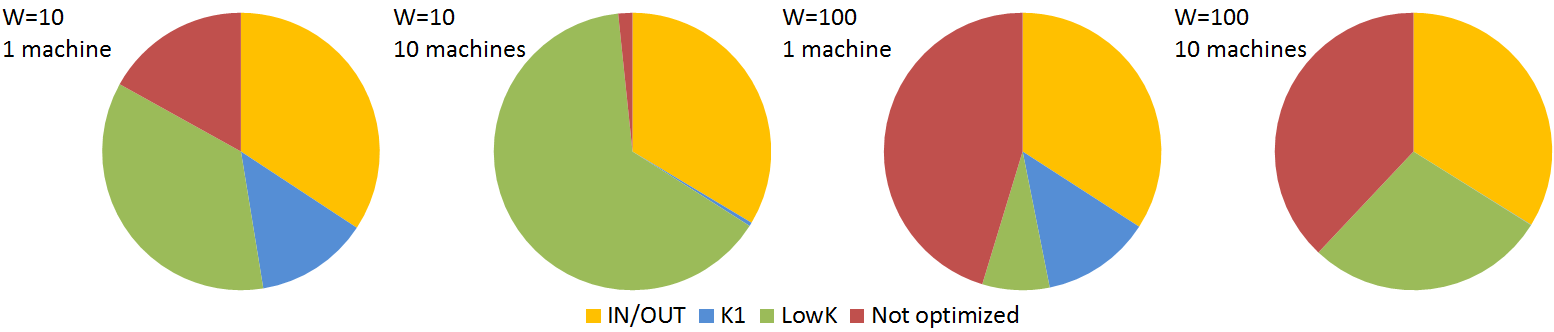}
\centering
\caption{Frequency of K-means optimizations trigger for data sets of 5000 points on various conditions}
\label{fig:opticharts}
\end{figure*}

We have tested our solution using the provided metadata in various configurations.
We summarize some of the most representative runs in Table~\ref{tab:runs}.
For the cases of 1000 machines we have used 12 worker threads and for the case of 1 machine a single worker thread.
The chosen window size was 10 in all three tests.
We have tested both an implementation using the final synchronization step in the solution architecture described in~\ref{sec:solution} as well a solution where anomalies detected by worker threads are simply pushed directly to the output queue.
We observed in various runs that the flavor we did not synchronize managed to perform relatively well in respect to output correctness (i.e., producing anomalies in the right order).
We believe the reason for this was the relatively low frequency when an anomaly appears. 
Also, an interesting aspect was that the solution flavor for 1000 machines that we did not synchronize performed just mildly better than the synchronized flavor. 
We can conclude therefore that our synchronization mechanism does not incur a very high overhead.   

\begin{table}
\begin{center}
 \begin{tabular}{||c c c c c||}
 \hline
 \thead{Points} & \thead{Machines} & \thead{Throughput \\ (MB/s)} & \thead{Latency \\ (ms)} & \thead{Sync} \\ [0.5ex]
 \hline\hline
 63 & 1000 & 36.7 & 6.79 & Yes \\
 \hline
 63 & 1000 & 36.1 & 6.45 & No \\
 \hline
 10000 & 1 & 31.6 & 5.68 & N/A \\ [1ex]
 \hline
\end{tabular}
\vspace{5pt}
\caption{Performance results - average for 5 runs on the evaluation platform}
 \label{tab:runs}
\vspace{-20pt}
\end{center}
\end{table}

We have also measured the frequency of triggering the optimizations implemented for the K-means processing stage described in Section~\ref{sec:optimizations}.
The results are illustrated in Figure~\ref{fig:opticharts}.
We have chosen specifically two cases of workloads provided by the Grand Challenge Competition organizers.
We have observed that testing for both a window size of 10 and 100, the optimizations were triggered for more than 50\% of the processed windows.
The optimizations are mutually exclusive with respect to a processed window in the following priority: \emph{IN/OUT}, \emph{K1}, \emph{LowK}.
The exact frequency of triggering each optimization for the total number of processed windows over 5000 points in the workloads is provided in Table~\ref{tab:opti}.

\begin{table}
\begin{center}
 \begin{tabular}{||c c c c c||}
 \hline
 \thead{Machines and \\ window size} & \thead{Total \\ windows} & \thead{IN/OUT} & \thead{K1} & \thead{LowK} \\ [0.5ex]
 \hline\hline
 1 / 10 & 274505 & 34.26\% & 13.16\% & 35.63\%  \\
 \hline
 10 / 10 & 2745050 & 33.58\% & 0.43\% & 64.32\% \\
 \hline
 1 / 100 & 269555 & 34.11\% & 12.74\% & 7.82\% \\
 \hline 
 10 / 100 & 2695550 & 33.86\% & 0.03\% & 28.11\% \\  [1ex]
 \hline
\end{tabular}
\vspace{5pt}
\caption{Exact frequency of each K-means optimization trigger}
 \label{tab:opti}
\vspace{-20pt}
\end{center}
\end{table}

In respect to the lazy optimization of the Markov modelling, we observed that the number of clusters for the sensors in the provided 1000-machines metadata was 53 on average.
This yields a 53x speed-up of the Markov modelling stage on average, as detailed in Section~\ref{sec:optimizations}.

During the preliminary tests, we have used a multi-node configuration composed of multiple Docker containers (from 2 to 7) using a baseline message passing.
The nodes were evenly spread across the available servers.
The evaluation revealed that the throughput, in this case, was not increased compared to a single node configuration, reaching about 40MB/s.
Considering on these results we decided to focus on a multi-threaded single node solution.

One further attempt to optimize our parsing stage was to use native C language solution.
We have observed that interfacing with a native implementation through JNI generated higher latencies than the original Java implementation, on average 2x time.
However, we still consider other options to explore for achieving this (e.g., sockets, shared memory).

\section{Discussion and potential enhancements}
\label{sec:discussion}

Although the evaluation platform offered a cluster of servers, the experiences presented in Section~\ref{sec:evaluation} led us to the conclusion that one multithreaded single node solution with highly optimized processing stages is a perfectly feasible approach in the given context.
We actually believe that a multiple node solution might suffer more from the overhead of passing messages over the network and additional synchronization than the gain offered by a distribution limited to 3 physical machines.
Also, a factor for choosing the one node solution was the amount of memory available on a single server (256GB), which we nevertheless carefully evaluated with respect to consumption (e.g., strictly considering the values in the window queues held in memory for 1000 machines, with an average of 55 stateful sensors per machine and a window size of 500, the occupied space would not exceed 300MB).

Besides the optimizations already implemented, we also considered some other potential enhancements of our solution.
One of them is maintaining the $K$ cluster centroids and the $W$ values within a window, both sorted ascending by value.
This would allow a significant improvement applicable to all iterations of the K-means stage, independent of the window content, dropping the theoretical running time-complexity from $O(K \times W)$ to $O(K+W)$.
In order implement this, first, the values within a window have to be kept sorted in some data-structure that would allow efficient insertions, deletions, and the iteration in increasing order.
A balanced tree set (e.g., AVL, Red-Black Tree) would allow the first two operations in logarithmic time and the latter in linear time.
The centroid values can be kept in a simple sorted array since only their values change from one round to another and there are no insertions or deletions after the initialization.
Using these structures in a K-means window processing would make possible to find the assigned cluster to one observation in $O(1)$ average time.
The algorithm would start by iterating all sensor values smaller than the average of the first two centroid values.
All these values are assigned to the first cluster.
Then, the following (higher) values are iterated and assigned to the second cluster, limited by the average of the next two centroid values.
The average is the boundary between clusters.
The process continues until all values are assigned.
The new cluster centroids can be updated in the same process to avoid further iterations.
The complexity analysis is very similar to the one of the merge-sort of two sorted arrays and would provide the $O(K+W)$ complexity.
This is possible because of the fact that the K-means is applied on 1-dimensional data.

Another potential optimization would be a memoization applied to the window contents, more precise by keeping a cache of the results obtained from previously processed similar windows.
This would require a fast enough hash algorithm for the window values and a low dispersion for the values emitted by each sensor.

However, an integration of these ideas and others would also increase the code complexity and probably also introduce needs for new synchronization points, so such extensions would require a careful evaluation.

\section*{Acknowledgements}
\begin{wrapfigure}{r}{0.10\textwidth}
\includegraphics[width=\linewidth]{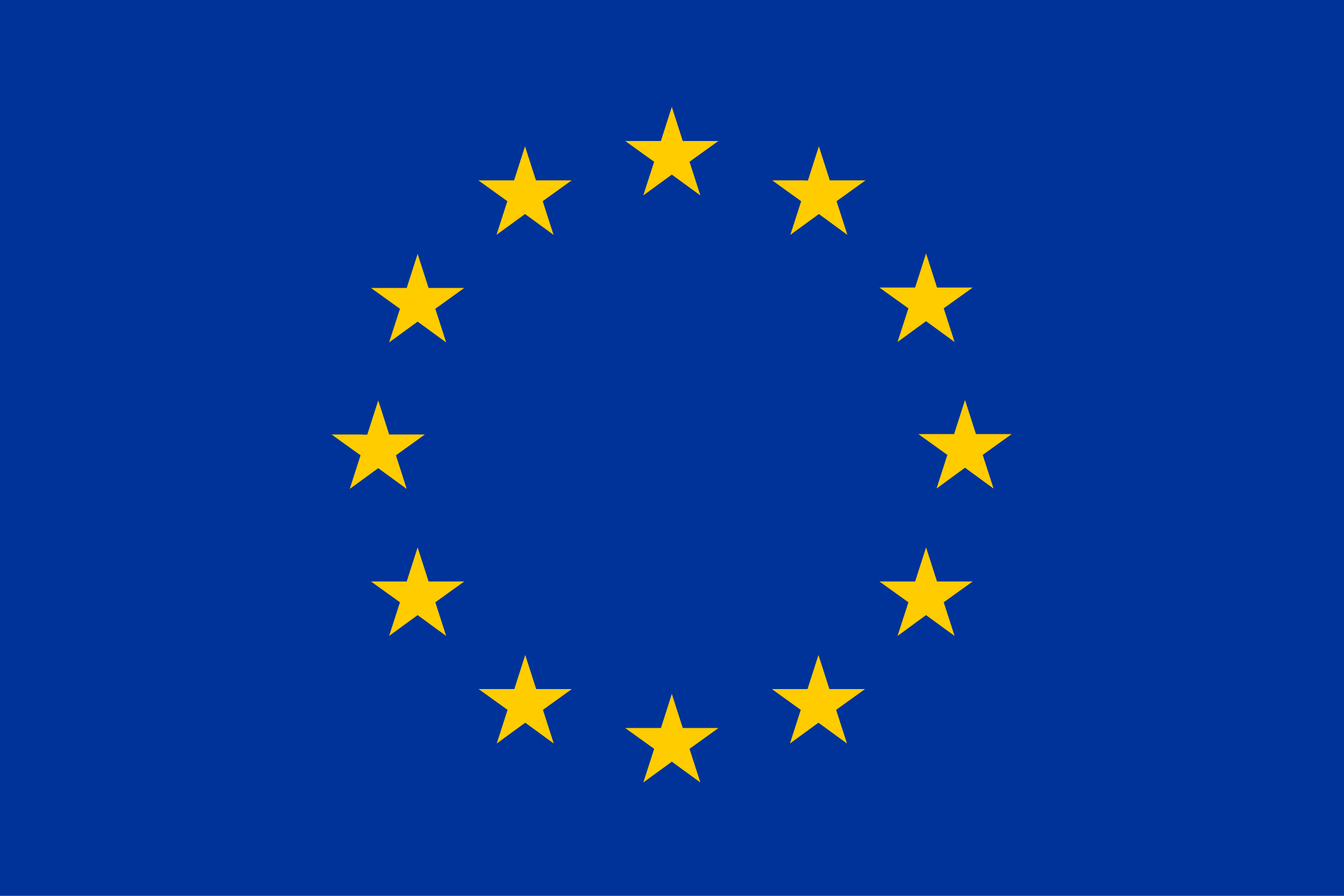}
\end{wrapfigure}
This work is partly funded from the \emph{European Union{\textquotesingle}s Horizon 2020 research and innovation programme} under grant agreement No 692178. This work was also partly supported by a grant of the Romanian National Authority for Scientific Research and Innovation, CNCS/CCCDI - UEFISCDI, project number 10/2016, within PNCDI III.

\bibliographystyle{ACM-Reference-Format}
\bibliography{sigproc}

\end{document}